\documentclass[twocolumn]{aastex62}
\usepackage{amsmath}
\hypersetup{linkcolor={red!90!black},citecolor={blue!60!black},filecolor=cyan,urlcolor={blue}}
\received{February 25, 2019}
\accepted{May 10, 2019 }

\shorttitle{Reconnection VHE Emission in SgrA*}
\shortauthors{Rodr\'iguez-Ram\'irez, de Gouveia Dal Pino \& Alves Batista.}

\begin{document}
\title{\textbf{VHE Emission from Magnetic Reconnection in the RIAF of SgrA*}}
\correspondingauthor{J. C. Rodr\'iguez-Ram\'irez}
\email{juan.rodriguez@iag.usp.br}
\author{Juan Carlos Rodr\'iguez-Ram\'irez}
\author{Elisabete M. de Gouveia Dal Pino}
\author{Rafael \surname{Alves Batista}}
\affiliation{
Instituto de Astronomia,  Geof\'isica e Ci\^{e}ncias Atmosf\'ericas (IAG-USP),
             Universidade de S\~{a}o Paulo.\\
             R. do Mat\~ao, 1226, 05508-090,  Cidade Universit\'aria, S\~ao Paulo-SP Brasil.
}

\begin{abstract}
The cosmic-ray (CR) accelerator at the galactic centre (GC) is not yet established by current observations. 
Here we investigate the radiative-inefficient accretion flow (RIAF) of Sagittarius A* (SgrA*) as a CR accelerator assuming acceleration by turbulent magnetic reconnection, and derive possible emission fluxes of CRs interacting within the RIAF (the central $\sim10^{13}$cm).
The target environment of the RIAF is modelled with numerical, general relativistic magneto-hydrodynamics (GRMHD) together with leptonic radiative transfer simulations.
The acceleration of the CRs is not computed here. 
Instead, we inject CRs constrained by the magnetic reconnection power of the accretion flow and compute the
emission/absorption of $\gamma$-rays due to these CRs interacting with the RIAF, through Monte Carlo simulations employing the {\tt CRPropa 3} code.
The resulting very-high-energy (VHE) fluxes are not expected to reproduce the point source HESS J1745-290 as the emission of this source is most likely produced at pc scales. 
The emission profiles derived here intend to trace the VHE signatures of the RIAF as a CR accelerator and
provide predictions for observations of the GC with improved angular resolution and differential flux sensitivity
as those of the forthcoming Cherenkov Telescope Array (CTA). 
Within the scenario presented here, we find that for mass accretion rates $\gtrsim10^{-7}$M$_\odot$yr$^{-1}$,
the RIAF of SgrA* produces VHE fluxes which are consistent with the H.E.S.S. upper limits for the GC and
potentially observable by the future CTA. 
The associated neutrino fluxes are negligible compared with the diffuse neutrino emission measured by the IceCube.
\end{abstract}

\keywords{
Accretion, accretion discs
--
Galaxy: centre
--
Magnetohydrodynamics (MHD)
--
Magnetic reconnection 
--
Radiation mechanisms: non-thermal
--
Astroparticle physics
}

\section{Introduction} \label{sec:intro}
The dynamic centre of our galaxy has been measured in the very-high-energy (VHE; $>$100 GeV) radiation regime 
by imaging atmospheric Cherenkov telescopes (IACTs) for over a decade 
\citep{2004ApJ...606L.115T, 2004ApJ...608L..97K, 2006ApJ...638L.101A, 2004A&A...425L..13A, 2009A&A...503..817A}. 
Particularly, the VHE point-like source J1745-290  measured by the High Energy Stereoscopic System (H.E.S.S.)  has been limited within an angular error of 13 arcsec \citep{2010MNRAS.402.1877A}
at the galactic centre (GC).
This limit encloses three plausible astrophysical counterparts 
within the central $\sim$10 pc:
a central spike of annihilating dark matter \citep{2012PhRvD..86j3506C, 2012PhRvD..86h3516B}, the pulsar
wind nebula (PWN) G359.95-0.04 
 \citep{2006MNRAS.367..937W},
and the supermassive black hole (SMBH) Sagittarius A* (SgrA*) 
\citep{2005ApJ...619..306A, 2005Ap&SS.300..255A, 2004ApJ...617L.123A,2017JCAP...04..037F}.

A hadronic scenario for the origin of J1745-290 has been favoured by measurements of the diffuse VHE
emission from the $\sim 200$ pc outskirts of the GC \citep{2006Natur.439..695A, 2016Natur.531..476H}. 
This diffuse emission is spatially correlated with the gas density of the central molecular zone (CMZ)
and its emission centroid coincides with the point-like source J1745-290. 
Such correlations then suggest that both the point-like as well as the diffuse fluxes might be produced
by proton-proton (p-p hereafter) interactions of cosmic-rays (CRs) with the gas around the GC.
Thus, if the origin of J1745-290 is mostly hadronic, the SMBH SgrA* could be powering CRs acceleration
up to PeV energies
\citep{2005ApJ...619..306A, 2005Ap&SS.300..255A, 2016Natur.531..476H}.

No variability in the flux of J1745-290 was found during the simultaneous H.E.S.S. and {\it Chandra} observations
performed in July 2005, when the X-ray flux of SgrA* increased by a factor of $\sim 9$ \citep{2008A&A...492L..25A}.
A subsequent analysis with increased statistics of data and improved analysis methods displayed no variability,
flaring, or quasi-periodic oscillations in the flux of J1745-290 \citep{2009A&A...503..817A}, and more recently,
a monitoring campaign of SgrA* with the MAGIC telescopes reported 
no significant variability of the emission from this source \citep{2017A&A...601A..33A}.
This lack of variability disfavours the origin of the bulk emission of the VHE point source within the
inner $\sim 100$ gravitational radii around the SMBH SgrA*, where simultaneous flares emitting at X-rays and
infrared (IR) bands are most likely produced 
\citep{2004A&A...427....1E, 2009ApJ...698..676D, 2011ASPC..439..309D,2016A&A...589A.116M, 2018arXiv181205764B, 2019arXiv190202933O}.

Another reason to not expect the emission of J1745-290 to be produced within few gravitational
radii is the following.
CRs leading to a flux compatible with J1745-290 from p-p interactions also produce secondary leptons.
If these hadronic interactions take place in a medium with a magnetic
field of $\sim$10 G or larger, as in the RIAF of SgrA*, the synchrotron emission of the secondary leptons may overshoot the 
quiescent X-ray flux from the GC\footnote{If the emission of J1745-290 is produced by p-p interactions at pc scales,
the secondary leptons diffuse in a medium with a considerably lower magnetic field (of the order of
$\sim$100 $\mu$G) and thus the synchrotron emission is lower, not conflicting with the quiescent
X-ray emission from the GC.},
as shown by \cite{2005ApJ...619..306A, 2005Ap&SS.300..255A}.

The unlikely origin of J1745-290 at few gravitational radii 
does not exclude however, the possibility of CR acceleration in the immediate vicinity of the SMBH.
CRs that escape their acceleration zone could produce the observed emission of J1745-290
further out, while diffusing within the interstellar medium at pc scales,
since given the angular resolution of the H.E.S.S. instrument, $\gamma$-rays produced within the central $\sim 10$ pc still appear as point-like emission. 
This scenario, originally proposed by \cite{2005Ap&SS.300..255A}, 
explains the spectrum of J1745-290 
and can be extended to also match the data of the  {\it Fermi} source 1FGL J1745.6-2900
\citep[e.g.,][]{2006ApJ...647.1099L, 2011ApJ...726...60C, 2012ApJ...753...41L}.

However, this interpretation as well as the current VHE observations from the GC give no direct information
about the specific sites of CR acceleration or the acceleration mechanism.
VHE fluxes observed with enhanced angular resolution and differential flux sensitivity like those potentially
obtained with the forthcoming CTA \citep{2013APh....43....3A, 2017arXiv170907997C} 
will be valuable to localise the presumed sites of CR acceleration at the GC.

In this work, we focus on possible VHE fluxes produced within tens of gravitational radii around 
the SMBH SgrA*. We particularly investigate the scenario of VHE emission produced by hadronic
interactions in the accretion flow, where these CRs are accelerated by turbulent magnetic
reconnection in the RIAF of SgrA*. 
In order to explore their detectability and how much they contribute to the current VHE emission
from the GC, we compare the derived emission profiles with the differential flux sensitivity of CTA,
as well as with the current emission of the source J1745-290.

The acceleration of CRs, and production of neutrinos and non-thermal radiation
has been previously discussed
in the framework of RIAFs and standard accrection disks considering  one-zone models
\citep[e.g.,][]{2010IJMPD..19..729D,2010A&A...518A...5D,2010A&A...519A.109R,2015MNRAS.449...34K,2015ApJ...806..159K,2016MNRAS.455..838K}.

Particle acceleration by magnetic reconnection has been studied
as an important emission process in BH systems, from BH binaries to active galactic nuclei 
\citep{2005A&A...441..845D, 2015ApJ...799L..20S,2015MNRAS.449...34K,2015ApJ...802..113K, 2016MNRAS.455..838K}.

The development of magnetic reconnection in accretion flows around BHs has been tested numerically by several authors 
\citep[e.g.,][]{
2008ApJ...682.1124K,
2014MNRAS.440.2185D,
2015MNRAS.446L..61P,
2018ApJ...864...52K,
2018arXiv180906742D,
2018ApJ...853..184B}.
In particular,  fast  reconnection occurrences  induced by the magneto-rotational-instability turbulence have been detected in multidimensional general relativistic (GR)
magneto-hydrodynamic (MHD) RIAF simulations 
(\citealt{2018arXiv180906742D}; \citealt{2019arXiv190404777K})
employing a  current-sheet-search  algorithm \citep{2018ApJ...864...52K}.
Also,  stochastic CR Fermi acceleration by turbulent magnetic reconnection has been extensively investigated numerically
(\citealt{2011ApJ...735..102K};
\citealt{2012PhRvL.108x1102K};
\citealt{2016PhPl...23e5708G}; 
\citealt{2016MNRAS.463.4331D};
\citealt{2018MNRAS.481.5687P};
\citealt{2019arXiv190308982D};
\citealt{2019MNRAS.482L..60W};
see also \citealt{2015ASSL..407..373D} for a review).  
Acceleration by reconnection has been also considered recently as
a triggering process to allow for further CR stochastic acceleration in MHD simulations of the MRI turbulence in hot accretion flows \citep{2019MNRAS.tmp..336K}. These authors found that stochastic acceleration is able to energise CRs up to 10 PeV.

In the context of SgrA*, magnetic reconnection has previously been invoked to model its observed IR and
X-rays flares \citep{2016ApJ...826...77B, 2017MNRAS.468.2552L}. 
All these aforementioned results suggest that efficient CR acceleration by magnetic reconnection may occur
and produce detectable VHE emission in the vicinity of the SMBH SgrA*.

In our approach, we assume that CRs are accelerated by magnetic reconnection and derive the implied VHE
fluxes constraining the total injection of CRs by the magnetic reconnection power of the system.
Such predicted VHE fluxes are not expected to coincide with the emission of the point source J1745-290, 
since as mentioned above, 
the nature of this source
is probably different from the emission produced
in the immediate vicinity of the SMBH.
Thus, the emission profiles derived here intend to be 
predictions for future observations with improved angular and energy sensitivity like those of the forthcoming CTA.

In the next section, we describe the numerical GRMHD
simulation and the GR radiative transfer calculations that we employ to model the accretion flow 
environment where CRs interact. The interactions and emission of CRs in the accretion flow are
simulated with a Monte Carlo approach that we describe in Section 3.
In Section 4, we describe and discuss different simulated models for the VHE fluxes associated to different values
of the mass accretion rate.
Finally, we present our conclusions in Section 5.

\section{The numerical GRMHD model for accretion plasma environment}
\label{sec:environment}

\begin{figure}
   \centering
   \includegraphics[width=\columnwidth]{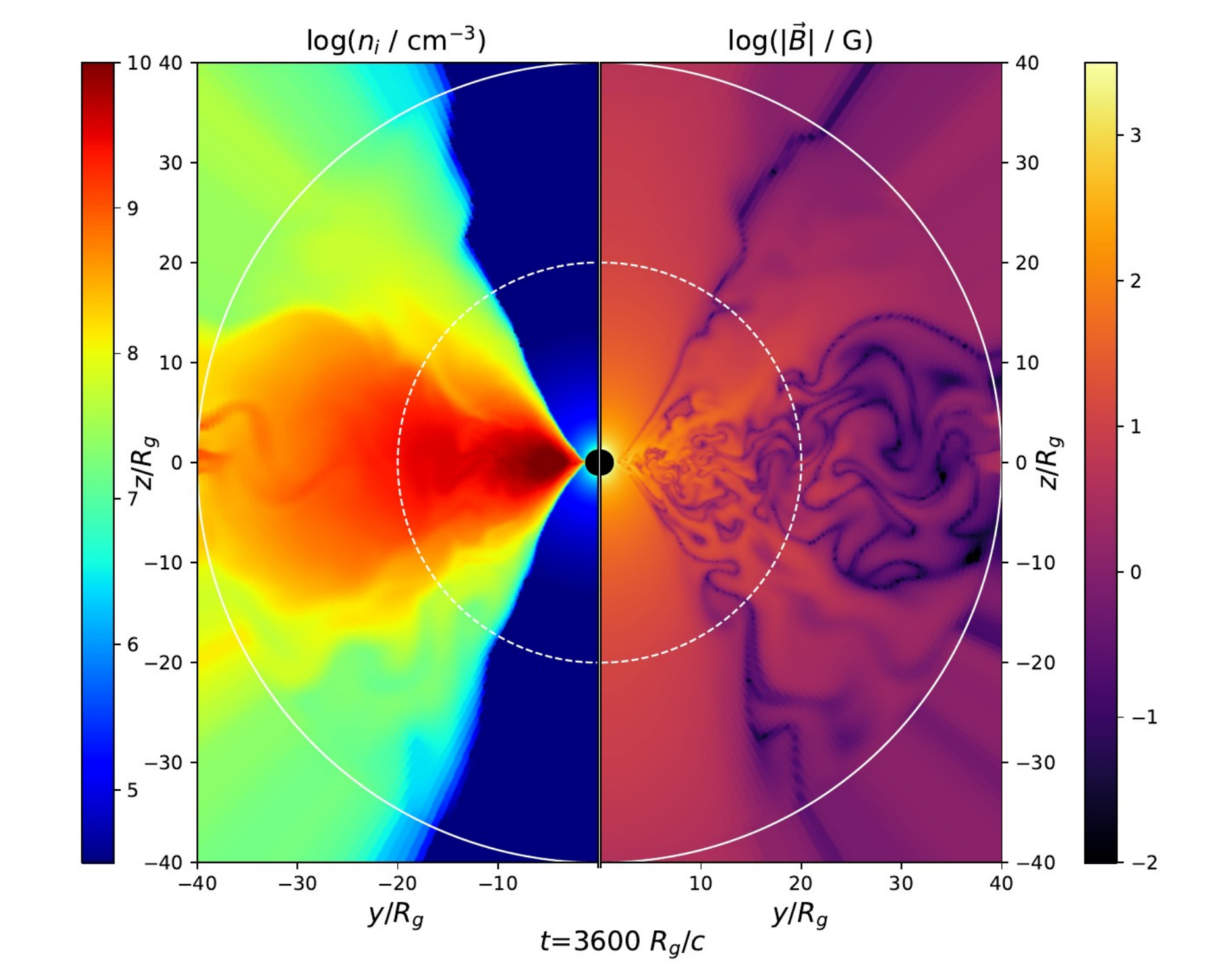}
      \caption{
Gas number density and magnetic field intensity maps corresponding to a snapshot at the time evolution
$t=3600 R_g/c$ of the accretion flow simulation described in Section 2, employing the axis-symmetric {\tt harm} 
code.
The inner white circle represents the spherical boundary of CR injection and the outer circle
the boundary for $\gamma$-ray detection in the CR simulations described in Section 3.
}
\label{FigrhoB}
\end{figure}

\begin{figure}
\centering
\includegraphics[width=\columnwidth]{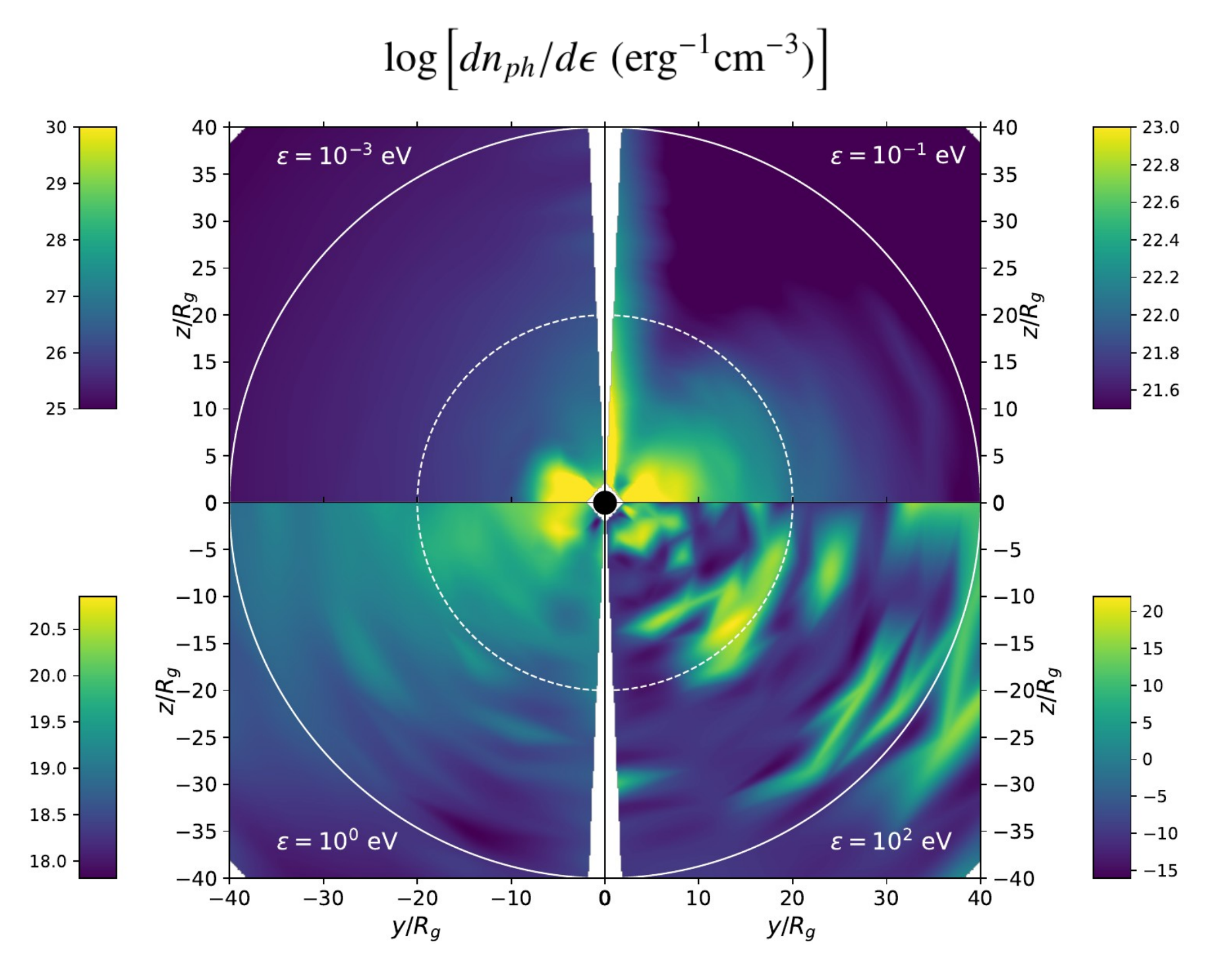}
\caption{ 
Photon field maps of synchrotron plus inverse Compton radiation corresponding to the accretion flow
snapshot of Fig.~\ref{FigrhoB}, calculated with the {\tt grmonty} code assuming the proton-to-electron temperature
ratio of $T_p/T_e=80$. The photon field is shown at different energy bands in each panel, as indicated.
 }
\label{Figdnde}
\end{figure}

Due to its bolometric luminosity and accretion rate constrained by observations,
the emitting material of SgrA* has been classified as a RIAF \citep{1995Natur.374..623N, 2003ApJ...598..301Y, 2014ARA&A..52..529Y}.
In this work we consider that CRs potentially accelerated by turbulent magnetic reconnection in the 
accretion flow
(e.g., \citealt{2015ApJ...799L..20S}) 
will first interact within the RIAF environment before escaping to the outer regions.
To model the gas density, magnetic and photon fields profiles of the plasma, we
employ the GRMHD axi-symmetric {\tt harm} code \citep{2003ApJ...589..444G, 2006ApJ...641..626N} together
with the GR radiative transfer {\tt grmonty} code \citep{2009ApJS..184..387D}.

We consider a numerical simulation where the accretion is triggered by magneto-rotational instability
on a torus initially in hydrostatic equilibrium \citep{1976ApJ...207..962F}, similar to the studies by e.g., 
\cite{2009ApJ...706..497M}, \cite{2018A&A...612A..34D}, \cite{2018MNRAS.478.1875J}, and other authors.
The initial torus of the simulation has an inner radius at $6 R_g$, a maximum gas 
pressure at $12R_g$, and is threaded with a poloidal magnetic field following the iso-density 
contours  with a minimum plasma beta\footnote{ $\beta$ gives the ratio between the thermal and the magnetic pressures of the plasma flow.} $\beta=50$. We use the dimensionless spin parameter $a=0.94$, the gas specific heat ratio $\gamma=4/3$, and a 
256x256 resolution for the $R$ and $\theta$ spherical coordinates.

The soft radiation field, $n_{ph}$, of the accretion flow is calculated with post-processing radiative
transfer (i.e., assuming  that radiation pressure is not important in the plasma dynamics) 
of synchrotron plus inverse Compton radiation. The emitting electrons are assumed to follow a relativistic 
Maxwellian distribution where their temperature is a fraction of the ions temperature.
The photon field is then obtained by calculating the radiation flux with the {\tt grmonty} code at different
radius and polar angles within the accretion flow zone.

The physical values of the gas density and the magnetic field profiles are obtained  as $\rho=\rho_0\rho'$, 
$B_i B^i=B_0^2 B'_i B'^i$, respectively, where $\rho'$ and $B'^i$ are the gas density and the 
3-magnetic  field in code units. In these 
code units, the gas density is normalised so that
$\rho'=1$  at the maximum pressure of the initial torus.

Given a fixed value for the density normalisation factor $\rho_0$,
the normalisation factor for the magnetic field is determined as $B_0=c\sqrt{4\pi\rho_0}$. This defines  the accretion flow characteristics in physical units for each  snapshot (Fig.~\ref{FigrhoB} depicts one of these snapshots from our simulation, see more below).

\begin{figure}
   \centering
   \includegraphics[width=\columnwidth]{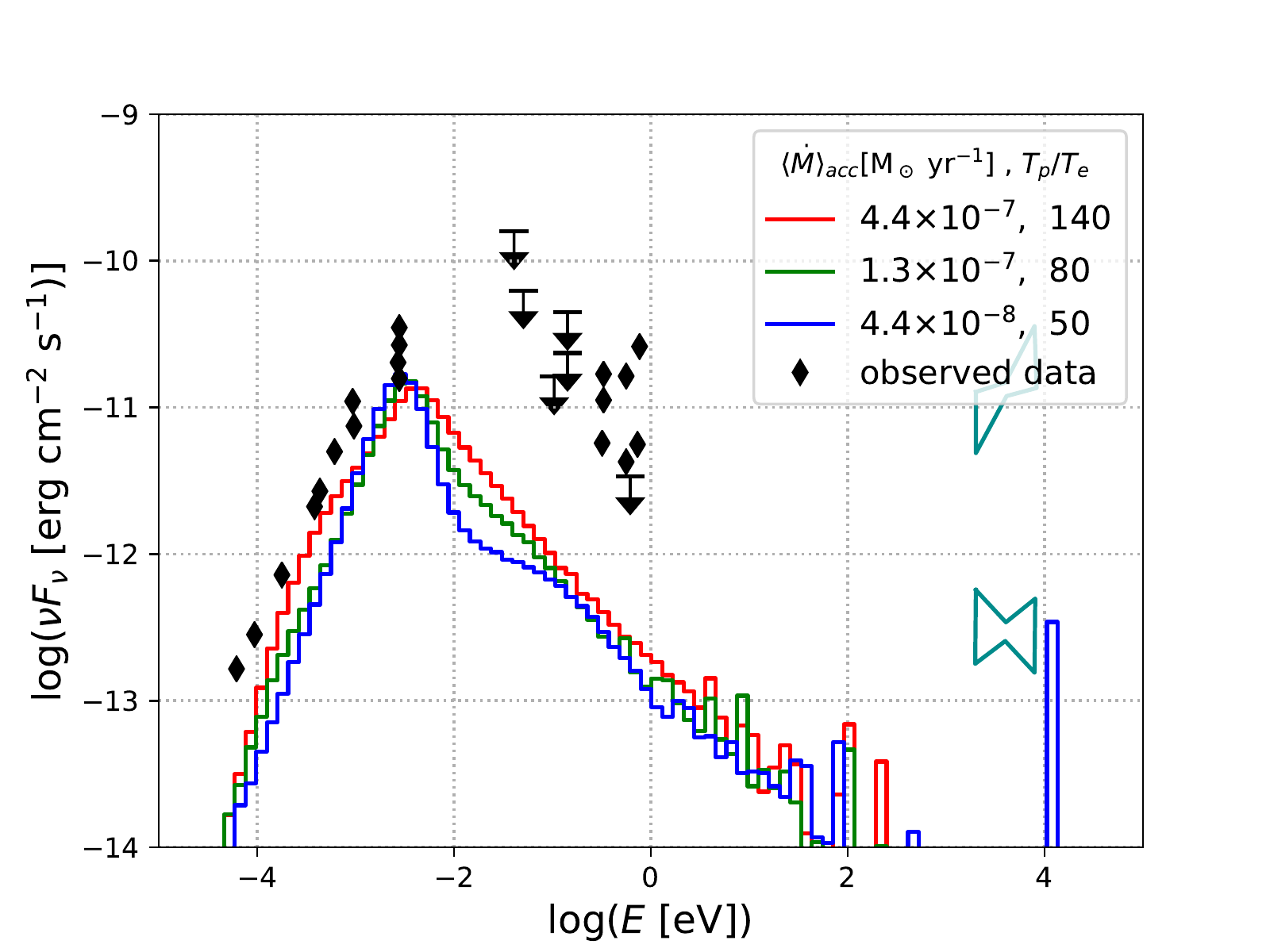}
\caption{
SED distribution due to synchrotron plus IC emission of the BH  accretion flow 
calculated with the {\tt grmonty} code
for three models with different mass accretion rates and proton to electron temperature ratio  (i.e. with different density normalisation). 
Each histogram gives the average leptonic SED obtained out of four snapshots of the accretion flow (see text). The data points
correspond to the observed SED of SgrA*, adapted from \cite{2009ApJ...706..497M}.
}
         \label{FigsoftSED}
   \end{figure}

The photon field density for each of these snapshots is obtained at the different points of the axi-symmetric flow  as:
\begin{equation}
\frac{ d n_{ph}}{d\epsilon}=
\frac{\nu L_\nu(R,\theta,\epsilon, T_p/T_e )}{4\pi c\epsilon^2 R^2}\,,
\end{equation}
where the differential luminosity $L_\nu$ is provided by the radiative transfer calculation at different
radii and polar angles (Fig.~\ref{Figdnde}). 
The value of the proton-to-electron temperature ratio, $T_p/T_e$, used to obtain the photon field of
a given snapshot is chosen such that the calculated spectral energy distribution (SED) due to synchrotron
plus IC is consistent with the observed radio to IR data of SgrA* (see Fig.~\ref{FigsoftSED}).

Summarising, the spatial profiles of $\rho$, $B^i$ and $n_{ph}$ (gas density, magnetic field, and soft
photon field) in the accretion flow are given by the GRMHD + GR radiative transfer simulations which
are normalised by arbitrarily choosing the gas density unit $\rho_0$ together with the constraint of the 
radio-IR data of SgrA* (to define the value of the ratio $T_p/T_e$).
Equivalently, we can freely choose the mass accretion rate\footnote{
The accretion rate is defined as $\dot{M}_{acc}=\int dr d\theta \,\sqrt{-g}\rho u^r$, where $g$ is the
determinant of the metric tensor and $u^r$ is the radial component of the fluid velocity.}
and then define the normalisation factors $\rho_0$, $B_0$, and $T_p/T_e$.

In Fig.~\ref{FigrhoB}, we show the gas number density and magnetic field maps corresponding to the snapshot  $t=3600\,R_g/c$ of the simulation as described in 
this section.
The physical units of the maps  in Fig.~\ref{FigrhoB} are obtained using the gas density normalisation factor
of $\rho_0=1.59\times 10^{10}$ cm$^{-3}$. 

In Fig.~\ref{Figdnde} we show the photon field maps associated to the snapshot of Fig.~\ref{FigrhoB}, assuming  $T_p/T_e=$80.

In the next section, we consider three different RIAF background models, each one obtained by choosing
a different gas density normalisation, as described.
For each background model, we calculate the emission of CRs injected in the accretion flow during a total time interval $\Delta t=320  R_g/c$ that encompasses  four subsequent snapshots of the simulation, namely  $t= 3440, 3520, 3600$,
and $3680\, R_g/c$, in order to mimic a continuous injection of CRs.  

In Fig.~\ref{FigsoftSED}, we show the  SEDs (averaged over these four snapshots) due to leptonic synchrotron plus IC emission for the three background models, each one labelled with  the corresponding mass accretion rates and temperature ratios.

\section{Monte Carlo simulation of cosmic-ray emission}
\label{sec:CRsim}

CRs accelerated in the RIAF of SgrA* by magnetic reconnection (see Section 4)
produce $\gamma$-ray fluxes due to their interactions with the
gas density, magnetic, and photon fields of the accretion flow environment. 
Whether these interactions are efficient enough to produce detectable $\gamma$-ray fluxes depends on the parameters of the accretion flow as well as on power and spectrum of the injected protons.
We investigate this potential hadronic $\gamma$-ray emission by simulating the propagation and interaction of 
CRs and their secondaries with an extended version of  the Monte Carlo code {\tt CRPropa 3} \citep{2016JCAP...05..038A}.
In these simulations, we assume that CRs are composed only by protons and we consider
(i) proton-proton interactions of CRs with the thermal background ions,
(ii) photon-pion interactions of CRs with the background soft photon field,
(iii) $\gamma$-$\gamma$  absorption of VHE photons by the background soft photon field with production of electron-positron pairs,
(iv) inverse Compton scattering of secondary leptons produced in the previous interactions, and
(v) synchrotron cooling to account for energy losses of charged 
particles.  Note that we do not track synchrotron photons as their energies are much lower than $\sim$\,1 GeV.

{\tt CRPropa} simulates 3D trajectories of charged particles by solving the relativistic Lorentz force equation in a flat space-time. 
In our approach, we take the magnetic field 3-vector $B^i$ given by the axi-symmetric
snapshots of the GRMHD simulation (see Section 2), rotate it to produce a magnetic field defined
in a 3D uniform grid, and use it in a flat spacetime without any further transformations to 
perform  the simulation of 3D CR propagation.
The error introduced by neglecting GR effects in the trajectory of the CRs and
photons in the resulting VHE emission decreases for $\gamma$-rays produced at larger distances from
the central BH (In Appendix B, we show the radial distribution for the production of VHE photons, for
chosen emission models derived in Section 4).

{\tt CRPropa 3} simulates the interactions of particles and photons along their trajectories using a Monte Carlo 
method and pre-loaded lookup tables for the interaction rates.
For our calculations, we implement into the code spatially-dependent interaction mean free paths (MFPs) 
calculated out of the gas density and photon field profiles of the GRMHD snapshots.
Thus, the MFP
for proton-proton interactions $\lambda_{pp}$, photo-hadronic interactions $\lambda_{p\gamma}$, photon-photon
pair creation $\lambda_{\gamma\gamma}$, and inverse Compton scattering $\lambda_{IC}$, are 
calculated  as: 
\begin{equation}
\lambda_{pp}^{-1}(E_p,R_i,\theta_j)=
\sigma_{pp}(E_p) n_{i}(R_i,\theta_j),
\label{lpp}
\end{equation}

\begin{equation}
\begin{split}
\lambda^{-1}_{p\gamma}(E_p,R_i,\theta_j)&=
\frac{m_p^2c^4}{2E_p^2}
\int_0^\infty d\epsilon\\
&\times
\frac{n_{ph}(\epsilon, R_i,\theta_j)}{\epsilon^2}
\int_{145 MeV}^{\frac{2 E_p \epsilon}{m_p c^2}}
d\epsilon' \epsilon' \sigma_{p\gamma}(\epsilon')\, ,
\label{lpg}
\end{split}
\end{equation}

\begin{equation}
\begin{split}
\lambda^{-1}_{\gamma\gamma}(E_\gamma, R_i,\theta_j)& =
\frac{1}{8E_\gamma^2}
\int_{2m_e^2c^4}^{4E_\gamma\epsilon_{max}} ds\\
& \times s \sigma_{\gamma\gamma}(s)
\int_0^\infty d\epsilon\frac{n_{ph}(\epsilon,R_i,\theta_j)}{\epsilon^2},
\label{lgg}
\end{split}
\end{equation}

\begin{equation}
\begin{split}
\lambda^{-1}_{IC}(E_e,& R_i,\theta_j)= 
\frac{1}{8 \beta E_e^2}
\int_{m_e^2c^4}^{m_e^2c^4+E_e\epsilon_{max}(1+\beta)} ds\\
&\times
\sigma_{IC}(s)(s-m_e^2c^4)
\int_0^\infty d\epsilon\frac{n_{ph}(\epsilon,R_i,\theta_j)}{\epsilon^2}.
\label{lIC}
\end{split}
\end{equation}

In equations (\ref{lpp})-(\ref{lIC}), $n_i(R_i,\theta_j)$ and $n_{ph}(R_i, \theta_j,\epsilon)$ are the gas 
number density and the photon field at discrete spatial points.
$E_p$, $E_e$, and $E_\gamma$ are the energies of protons, leptons, and $\gamma$-rays, respectively, and $m_p$ and $m_e$ the proton and electron mass, respectively.
$\sigma_{pp}$, $\sigma_{p\gamma}$,
$\sigma_{\gamma\gamma}$, and $\sigma_{IC}$ are the cross sections for the proton-proton interaction, photopion production,
photon-photon pair production and inverse Compton scattering, respectively.
In the integral of equation (\ref{lIC}) $\beta=\sqrt{1-m_e^2c^4/E_e^2}$.

Because {\tt CRPropa} does not perform proton-proton interactions, we have implemented it using the parametrisation for $\sigma_{pp}$ from \citet{2014PhRvD..90l3014K}. The fraction of the parent's energy taken by secondary particles is estimated using parametrisations from \citet{2006PhRvD..74c4018K}.
The expressions for the cross sections $\sigma_{\gamma\gamma}$ and $\sigma_{IC}$ can be found in
\cite{2015cwa} (see also \citealt{2012CoPhC.183.1036K} and \citealt{1998PhRvD..58d3004L}). 
The MFPs of Eqs. (\ref{lpg})-(\ref{lIC}) are calculated employing the {\tt CRPropa} 
tools\footnote{ \url{https://github.com/CRPropa/CRPropa3-data}},
that we modify in order to include the information of the gas density $n_i(R_i,\theta_j)$ and the photon field 
$n_{ph}(\epsilon,R_i,\theta_j)$.

 \begin{figure}
   \centering
   \includegraphics[width=\columnwidth]{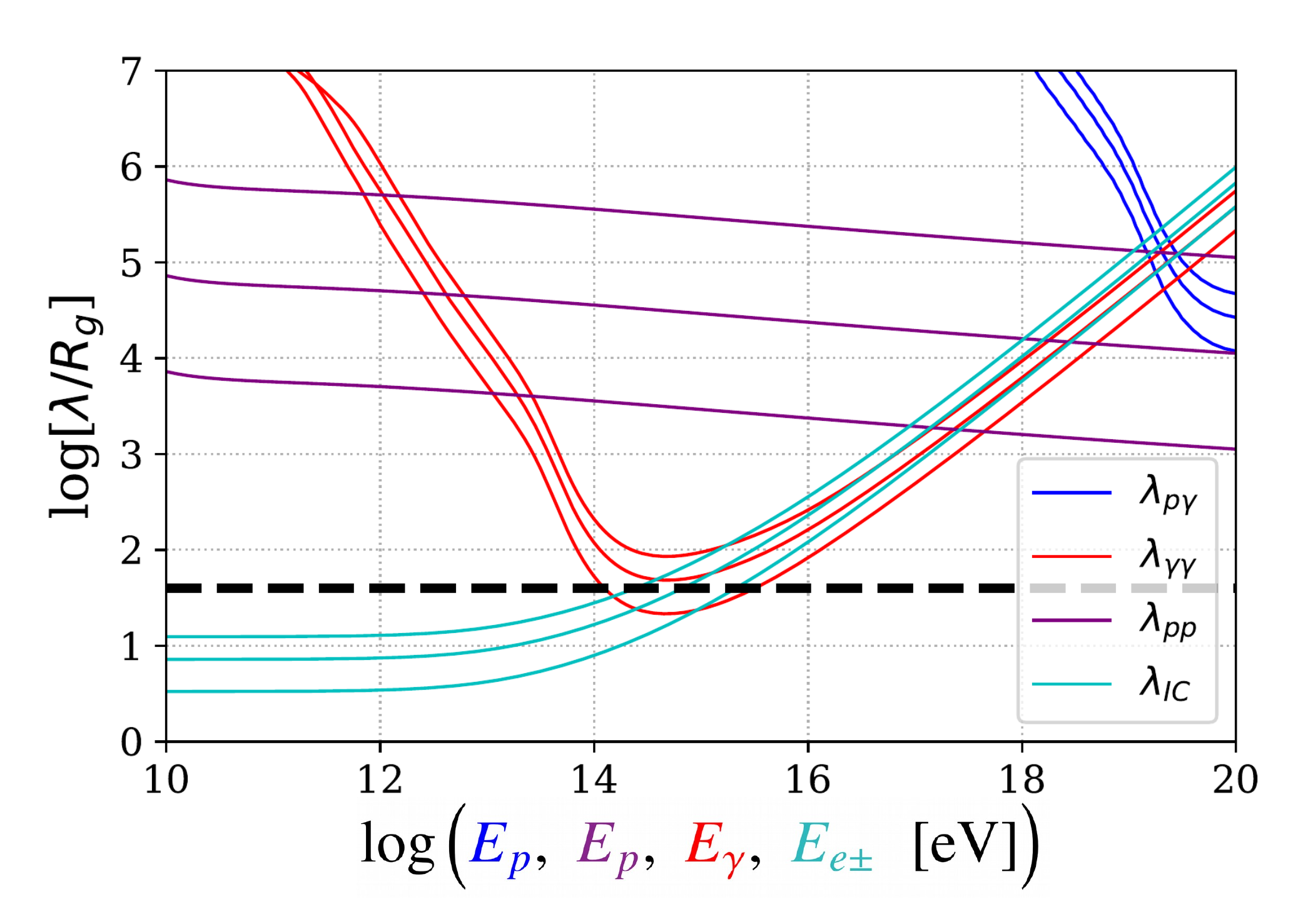}
      \caption{
Interaction mean free paths for particles and photons, in units of the gravitational radius of SgrA*.
The size of the accretion flow zone ($40R_g$) is indicated by the thick, dashed, black line. The 
interactions are calculated with the gas density and photon field obtained from the numerical GRMHD 
snapshot displayed in Figs. 1 and 2.
For the photon-pion ($\lambda_{p\gamma}$), pair production ($\lambda_{\gamma\gamma}$), 
and inverse Compton ($\lambda_{IC}$) interactions, the different curves are
calculated at the $\theta=45^{\mbox{\tiny{o}}}$ polar angle, and at 20 (lower-most lines), 30 (middle),
and 40 $R_g$ (upper-most) from the central BH. 
For the proton-proton interaction ($\lambda_{pp}$)
the different curves are calculated with the maximum gas density (lower-most), and with 0.1
(middle), and 0.01 (upper-most) fractions of the maximum gas density. 
              }
         \label{Figmfps}
   \end{figure}

In Fig.~\ref{Figmfps}, we show these interaction MFPs as a function of the energy, calculated
with the gas density and the photon field of the GRMHD snapshot of Figs. (1)-(2).
Although in the next section we consider different accretion flow backgrounds where the associated 
MPF curves are slightly different, the example shown in Fig.~\ref{Figmfps} qualitatively describes the 
overall behaviour of the interactions for all the background models considered in this work.
We see that protons with energies $\lesssim 10^{15}$ eV produce $\gamma$-rays principally due to proton-proton
interactions; $\gamma$-ray absorption by pair creation is significant for $\gamma$-rays with energies
within $\sim 10^{13}$-$10^{18}$ eV; and $\gamma$-rays produced due to IC scattering are mainly produced
by secondary leptons with energies $\lesssim10^{14}$ eV.
We note that $\gamma$-rays with energies $\lesssim10^{13}$ eV readily escape the RIAF of SgrA*, as previously
discussed by \cite{2005ApJ...619..306A}.
The production of $\gamma$-rays due to photo-hadronic processes within the RIAF is significant for protons
with energies $\gtrsim 10^{18}$ (ultra-high-energy cosmic-rays). 
However, in all models of this work we assume that the energies of the accelerated CRs are lower than 
 $\sim 10^{15}$ eV (see Section 4), and thus, photo-hadronic interactions
 are not expected to significantly contribute to the production of $\gamma$-rays. 

To calculate the emission due to a continuous injection of CRs, we employ four different snapshots with 80
$R_g/c$ time separation between each other, and simulate a burst-like propagation of CRs in each one.
We inject in each burst CRs with a power-law energy distribution with power-law index $\kappa$, and
exponential energy cut-off $\epsilon_{cut}$: 
\begin{equation}
\frac{dN_{CR}}{d\epsilon}\propto \epsilon^{-\kappa} \exp\{-\epsilon/\epsilon_{cut}\}.
\end{equation}
The energy of injected CRs in each snapshot is $E=W_{CR}\Delta t /4$, where $\Delta t=320 R_g/c$
is the total injection time interval and $W_{CR}$ is the CR injection power.\footnote{ We note that the resulting fluxes we obtain are approximately the same if we take only one snapshot and inject all the CRs energy in this single one. This is because we chose snapshots in which the system is already nearly in steady state, so that the particles interact essentially with same background in the four snapshots.}  

The simulations are performed with $10^6$ protons, with initial position and momentum isotropically 
distributed within a sphere of $20 R_g$ (represented as the inner dashed circle in Figs. 1-2) and we track photons
and particles until they attain a minimum energy of $10^{11}$eV, or they complete a trajectory length
of $160R_g$ (initiated from the parent protons), or else they cross a spherical detection boundary at
 $40R_g$ (represented as the outer circle in Figs. 1-2) from the central BH.
In Appendix A we describe in detail how we calculate the SED of the $\gamma$-rays from the 
output data of the simulations.
In the next section we describe the VHE emission models obtained from CRs interacting in different
background models, where the CR injection is constrained to the available magnetic reconnection
power of the background accretion flow.

\section{VHE fluxes from the RIAF of SgrA*}
\label{sec:VHEfluxes}

\begin{table*}
\caption{
Parameters of the emission profiles plotted in Fig.~\ref{VHE}.
}              
\label{table1}      
\centering                                      
\begin{tabular}{c c c c c c c}          
\hline\hline  \\                      
Model 
& $\langle \dot{M}_{acc}\rangle \times 10^{-7}$ [M$_\odot$ yr$^{-1}$] 
& $T_p/T_e$ 
& $W_{CR} \times 10^{37}$[erg s$^{-1}$]
& $W_{CR}/W_{rec}$
&$\kappa$ 
& $\epsilon_{cut}$ [PeV] \\
\hline                                   
 m$_{11}$  & $4.452$   & 140    &  30  & 0.8   & 2.4   & 0.05   \\
 m$_{12}$  & $4.452$   & 140    &  2   & 0.05   & 1.8   & 0.5    \\
 m$_{13}$  & $4.452$   & 140    &  0.8 & 0.02   & 1.3   & 0.5    \\

 m$_{21}$  & $ 1.335$  & 80     & 6.5  & 1.0    &  1.8   & 0.05   \\
 m$_{22}$  & $ 1.335$  & 80     & 6    & 0.92  &  1.8   & 0.5    \\
 m$_{23}$  & $ 1.335$  & 80     & 3    & 0.46  &  1.3   & 0.5    \\

 m$_{31}$  & $0.445$   & 50     &  1.3  & 0.96   & 1.0   & 0.05   \\
 m$_{32}$  & $0.445$   & 50     &  1.3  & 0.96   & 1.0   & 0.5    \\

\hline                                             
\end{tabular}
\end{table*}

As described, we consider three background RIAF models and calculate their VHE fluxes. Each background model
(i.e. the gas density and the magnetic and photon fields of the accretion flow)
is defined by choosing the value of the gas density normalisation, $\rho_0$, for the snapshots of the 
GRMHD simulation, and fixing the proton-to-electron temperature ratio, $T_p/T_e$, to obtain the photon
field which is constrained by the radio-IR data of SgrA* (see Section 2).
The magnetic reconnection power of a particular background model to accelerate the CRs, is obtained following the analytic
model for magnetic reconnection in magnetically-dominated accretion flows (MDAFs) of \cite{2015ApJ...799L..20S} 
(see also  \citealt{2005A&A...441..845D,  2015ApJ...802..113K}):
\begin{equation}
W_{rec}=
1.52 \times 10^{42} f
\left(\frac{\langle \dot{M}_{acc}\rangle}{\mbox{M}_\odot\mbox{yr}^{-1}}\right)
\left( \frac{T_p}{T_e}\right) \,  \mbox{erg s}^{-1},
\label{Wrec}
\end{equation}
where $\langle \dot{M}_{acc}\rangle$ is the average accretion rate over the four snapshots employed for
the calculation of the CR emission (see Sections 2-3 and Appendix A), the factor $f$ is
a combination of dimensionless parameters\footnote{$f\equiv A\Gamma^{-1}(11.5\alpha^{10/3}+14.89)^{1/2}$,
being $A$ the ratio of the height to the radius of the magnetic reconnection zone, $\Gamma^{-1}=[1+(v_A/c)^2]^{1/2}$
the  relativistic correction factor of the Alfv\'en velocity, and $\alpha$ the viscosity parameter. 
See \cite{2015ApJ...799L..20S} for details.} 
and here we adopt a fiducial value $f=4$.
The basic assumption of this work is that the energy of the CRs accelerated in the RIAF is provided by magnetic reconnection, as described in Section 1.
Thus, to calculate the VHE emission in the background models, we constrain the
power of CR injection, $W_{CR}$,  with the condition $W_{CR}/W_{rec}<1$.

\begin{figure}
   \centering
   \includegraphics[width=\columnwidth]{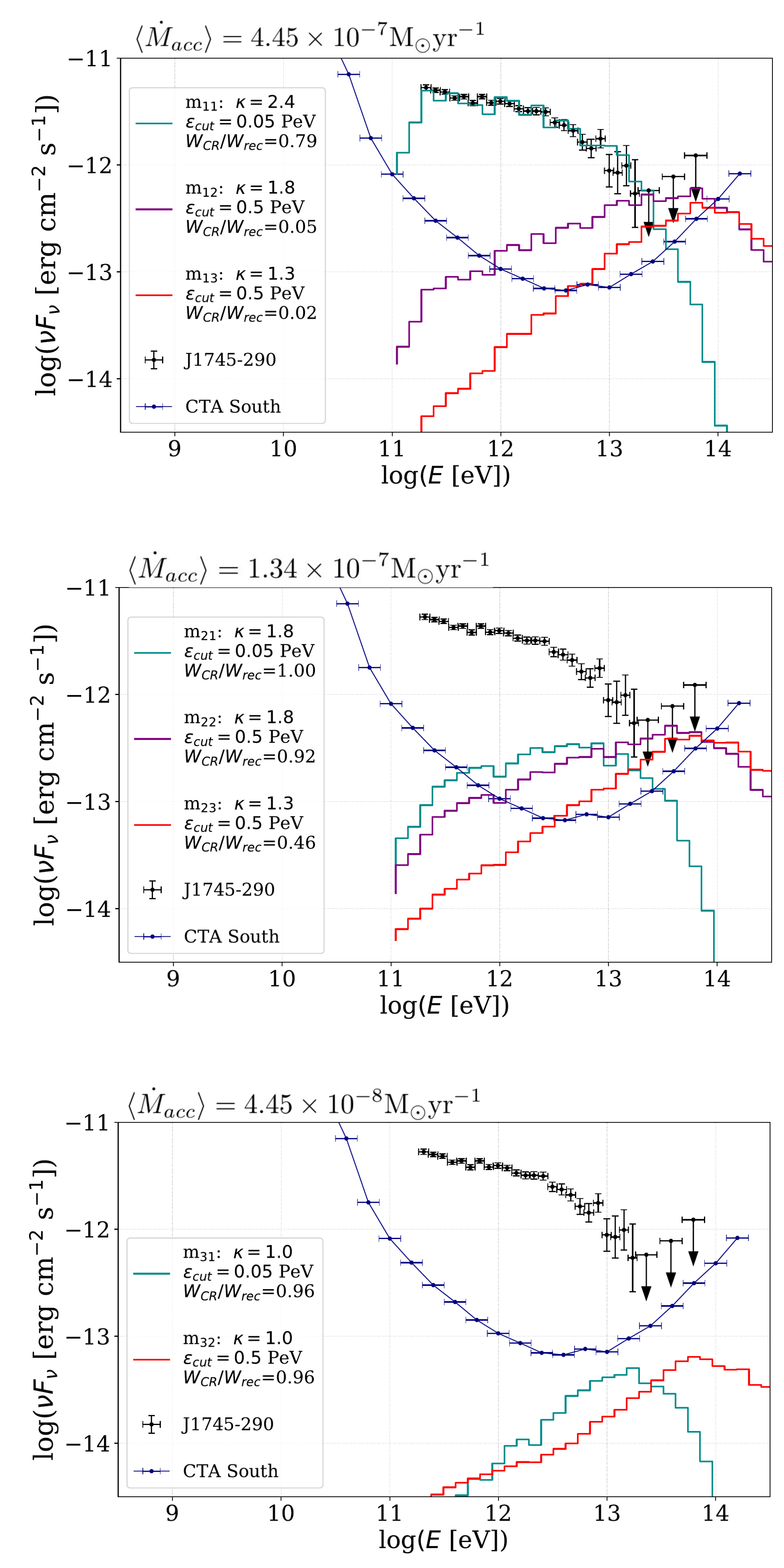}
      \caption{
SED of hadronic emission produced
by CRs accelerated by magnetic reconnection 
within the RIAF of SgrA* and calculated with the {\tt CRPropa 3} code
(see the text). Additional parameters of the emission profiles can be found in Table~\ref{table1}. 
In each panel we over-plot the data points of the source HESS J1745-290, adapted from \cite{2016Natur.531..476H},
and the differential flux sensitivity of the forthcoming CTA, adapted from \cite{2017arXiv170907997C}
}
         \label{VHE}
\end{figure}

Fig.~\ref{VHE} shows the calculated SED of the accretion flows with different accretion rates, and
the parameters  obtained for the plotted emission profiles
 are listed in Table~\ref{table1}. 
 
The emission curves in the upper panel of Fig.~\ref{VHE} correspond
to a background accretion flow with $\langle \dot{M}_{acc}\rangle=4.452 \times 10^{-7}$ M$_\odot$ yr$^{-1}$. 
The emission profile of model m$_{11}$  in this diagram is obtained injecting CRs with a power-law index $\kappa=2.4$
and  exponential cut-off energy $\epsilon_{cut}=5\times 10^{13}$ eV. 
This emission profile matches quite well the VHE data of the point source J1745-290.
However, as we mention in the introduction, such a VHE flux produced out of 
proton-proton interactions also implies an X-ray flux
(due to the synchrotron emission of the secondary leptons produced through the channel $p$+$p\rightarrow\pi^{\pm}\rightarrow$ e$^{\pm}$) that overshoots by $\sim$ one order of magnitude the quiescent X-ray emission of SgrA* \citep{2005Ap&SS.300..255A,2005ApJ...619..306A}.
This overestimation would not be conflicting if the flux of J1745-290 corresponded to flare episodes coincident
with those measured at X-rays energies from SgrA*. However, as mentioned in the introduction, variability,
flaring, or QPOs have not been found in the flux of the source J1745-290 
\citep{2008A&A...492L..25A, 2009A&A...503..817A}\footnote{Though one cannot disregard completely this model m$_{11}$, since higher sensitivity observations by forthcoming instruments like CTA, for instance, may be able to capture variability patterns in the VHE flux.}.
In contrast, the emission profile of the model m$_{12}$ obtained with CRs injected with the power-law
index $\kappa=1.8$ and cut-off energy $\epsilon_{cut}=0.5$ PeV produces a flux within the RIAF of SgrA*
that matches the  upper limits detected by H.E.S.S. in the VHE tail.
This emission profile peaks at $\sim 10^{13.5}$ eV with a flux one order of magnitude lower than the 
peak of the model m$_{11}$ and thus does not overshoot the  quiescent emission at X-ray energies.
Also, it requires $\sim$5\% of the reconnection power, and is above the differential flux sensitivity of
the future CTA in the energy range of $10^{12-14}$ eV.
The emission profile of the model m$_{13}$ corresponds to CRs injected with the power-law index $\kappa=1.3$
and cut-off energy $\epsilon_{cut}=0.5$ PeV. This emission model requires $2\%$ of the reconnection
power and is also above the CTA sensitivity.

The emission curves in the middle panel of Fig.~\ref{VHE} correspond to CRs emitting within a background with
 accretion rate of $\langle \dot{M}_{acc} \rangle=1.33\times 10^{-7}$ M$_\odot$ yr$^{-1}$.
In this case, the magnetic reconnection power is not sufficient to reproduce the bulk emission
of the source J1745-290 with any configuration of the parameters $W_{CR}$, $\kappa$, and $\epsilon_{cut}$.
The curve from model m$_{21}$ is produced with CRs injected with the power-law index $\kappa=1.8$
and cut-off energy $\epsilon_{cut}=5\times 10^{13}$ eV, and  peaks at $\sim 10^{12.5}$ eV with a flux of
$10^{-12.5}$ erg cm$^{-2}$ s$^{-1}$.
This flux profile makes no substantial contribution to the current data of the source J1745-290, but is, in principle, detectable by the future CTA.
The flux models m$_{22}$ and m$_{23}$ correspond to CR injection with the same values of $\kappa$
and $\epsilon_{cut}$ of the models m$_{12}$ and m$_{13}$, respectively, and they are also 
qualitatively similar to their counterparts.
They differ in that the models m$_{22}$  and m$_{23}$ require a larger 
fraction of the available magnetic reconnection power,
which is naturally expected since the rate of proton-proton interactions as well as the magnetic reconnection
power diminish for lower accretion rate (see Eq.~\ref{Wrec}).

Finally, it is shown in the lower panel of Fig.~\ref{VHE} that the VHE fluxes produced by RIAFs with
accretion rates of $\langle \dot{M}_{acc}\rangle=4.45 \times 10^{-8}$ M$_\odot$ yr$^{-1}$ and lower
are not detectable. 

Thus, based in the flux profiles of Fig.~\ref{VHE}, we conclude that magnetic reconnection
in the RIAF of SgrA* powers VHE fluxes that can be potentially detected by CTA, provided
that the mass accretion rate is $\langle \dot{M}_{acc}\rangle \gtrsim  10^{-7}$ M$_\odot$ yr$^{-1}$. 
With regard to the expected fraction of the magnetic reconnection power that should be available for CR acceleration, laboratory experiments of reconnection \citep{2010RvMP...82..603Y} and PIC simulations 
(e.g., \citealt{2017ApJ...843L..27W}) predict that  30\% to 50\% goes into particle acceleration and the remaining goes to heat the plasma. In this sense,  models m$_{12}$, m$_{13}$, and m$_{23}$ would be more favorable solutions\footnote{
This is also consistent with earlier studies of RIAF accretion flows which predict that a signiﬁcant 
fraction of the magnetic power has to be also
employed to heat the gas. Otherwise, the RIAF cannot keep its dynamical structure due to the low pressure, resulting in the formation of a geometrically thin, cold disc \citep{2014ApJ...791..100K}.
}.

\begin{figure}
   \centering
   \includegraphics[width=\columnwidth]{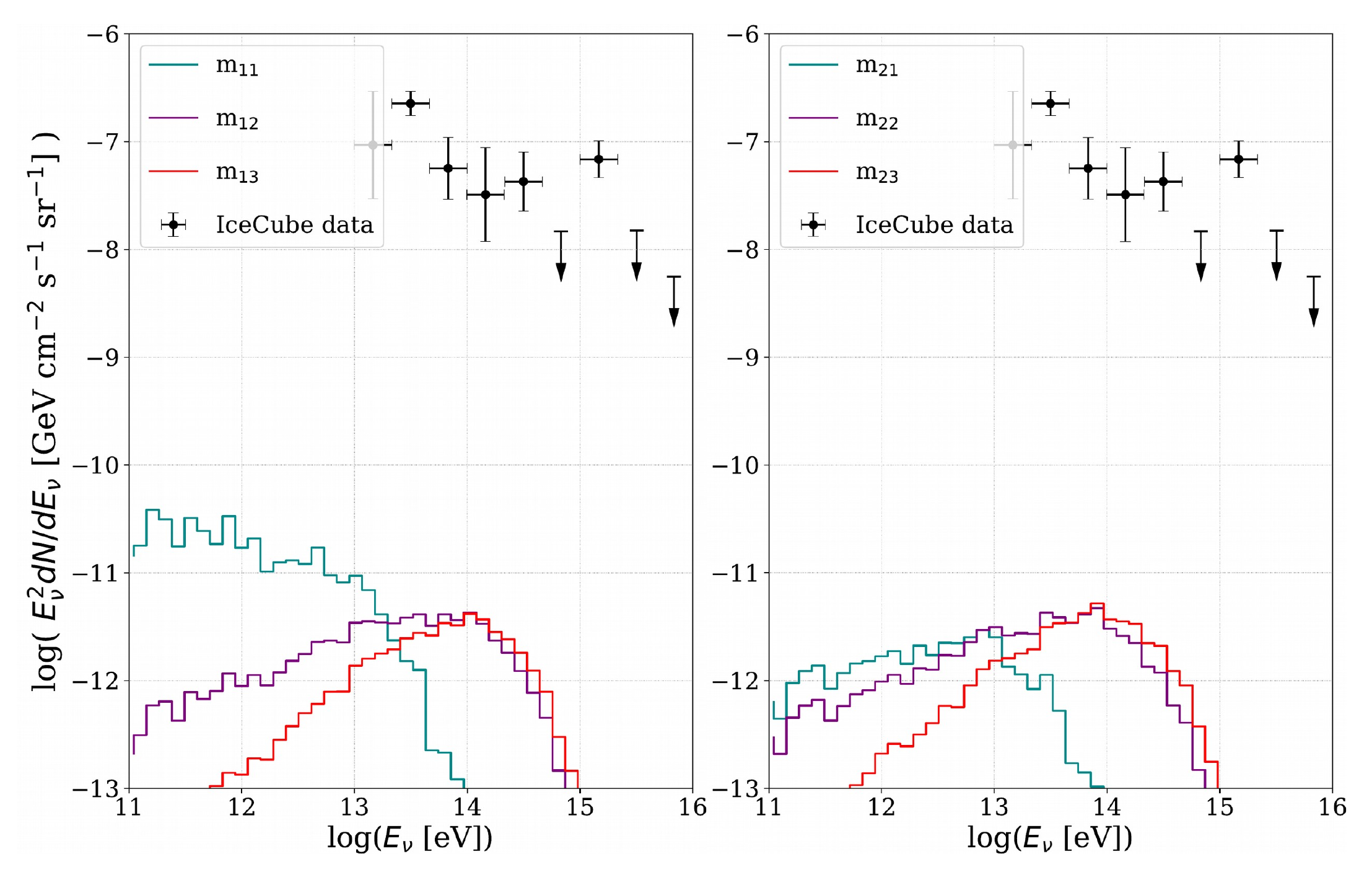}
      \caption{
      All-flavour neutrino fluxes associated to the emission profiles of the upper and middle  panels of Fig 5. The IceCube high-energy starting events are taken from \cite{2015ApJ...809...98A}.
}
         \label{nusSED}
\end{figure}

In Fig.~\ref{nusSED}, we show the neutrino fluxes associated to the emission models m$_{11}$, m$_{12}$, m$_{13}$, m$_{21}$, m$_{22}$ and m$_{23}$ of Fig. 5.
We see that the neutrino emission produced within the accretion flow of SgrA* is practically negligible compared to the IceCube diffuse neutrino flux.

As mentioned in Section 3, the error introduced by using the background environment of the GRMHD simulation in a flat spacetime and neglecting the GR curvature effects on the propagation of CRs and VHE photons, is lower for $\gamma$-rays produced at larger distances from the central BH. 
In Appendix B, we show that 
photons with energies $\gtrsim10$ TeV, are more uniformly sourced within
the 40$R_g$ volume than photons with lower energies. This is because
lower energy photons are produced by lower energy CRs which have smaller
Larmor radii and then are more easily trapped close to the BH where the magnetic field is more intense.
Thus, we conclude that the error of neglecting the GR effects in the calculation of the VHE emission
is lower for the emission models m$_{12}$, m$_{13}$, m$_{22}$, and m$_{23}$,
which peak at energies $>10$ TeV.
A more consistent approach considering the GR curvature effects on the propagation
of CRs and VHE photons (possibly by following the approach of \citealt{2019ApJS..240...40B}) is left for future
work.

\begin{figure}
   \centering
   \includegraphics[width=\columnwidth]{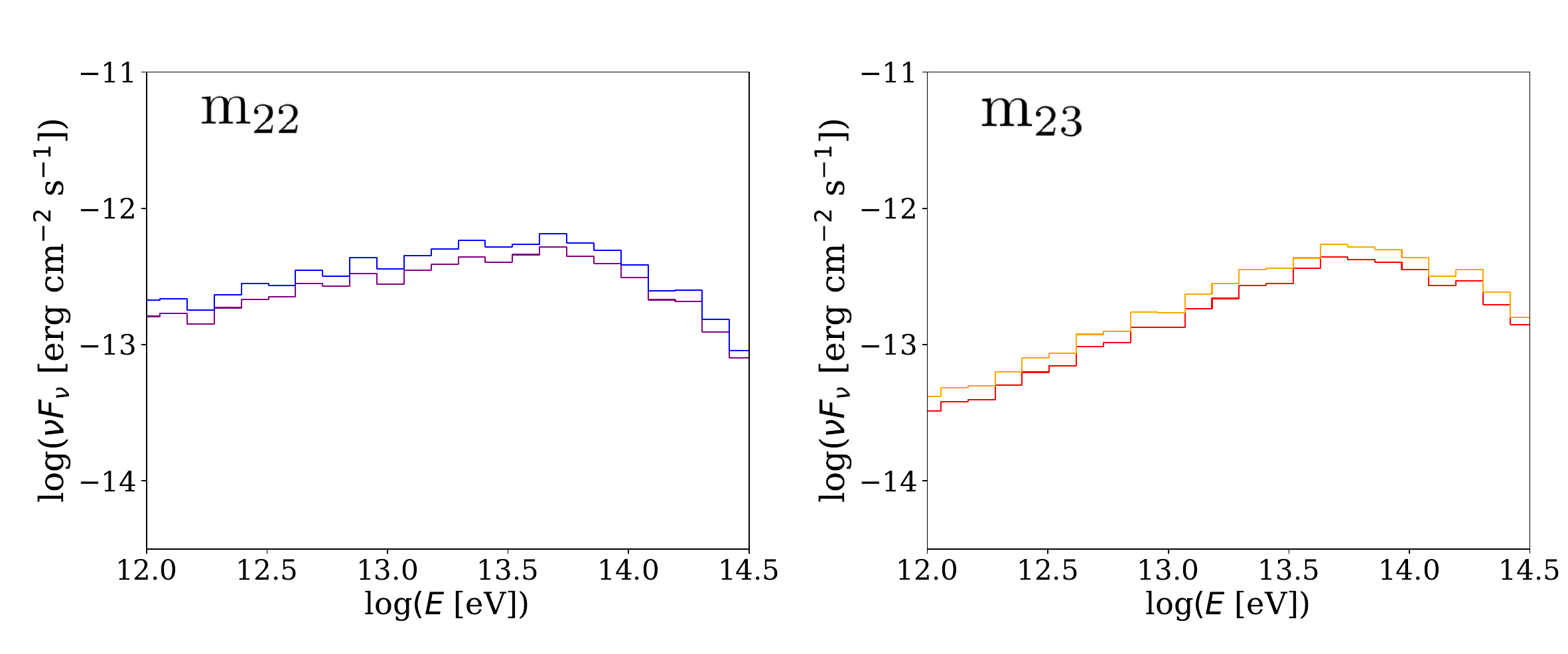}
      \caption{
Calculated SEDs corresponding to the models m$_{22}$ and m$_{23}$ (see Table 1)
for two different CR injection configurations. 
The purple and red curves correspond to CR injection homogeneously distributed within a
sphere of $20 R_g$ (as in Fig.~\ref{VHE}). The blue and orange curves correspond to CR 
injection with the same model parameters, but weighting the previous simulation output to 
limit the CR injection within the volume defined by the polar angle range
$\theta \in[\pi/4, 3\pi/4]$ 
and the spherical radius range $r\in$[2$R_g$, 20$R_g$].
}
         \label{comparison}
\end{figure}

In axy-symmetric numerical GRMHD RIAFs, strong magnetic reconnection events are localised  mostly
within the turbulent torus and in the interfaces between inflows and outflows, where several magnetic 
field reversals are identified, as can be seen in \citep{2018arXiv180906742D} and \citep{2019arXiv190404777K},
where a fast magnetic reconnection tracking algorithm \citep{2018ApJ...864...52K} is employed. 
Because in this work we do not analyse the details of the magnetic reconnection sites produced 
by the simulated accretion flow, we injected isotropically the accelerated CRs within a sphere of 20$R_g$, 
where the magnetic fields take the highest values (see Section 3).
If alternatively the CR injection is limited to the region of the
accreting torus, the flux of the VHE emission is increased
by a factor of $\sim$2, as can be seen in Fig.~\ref{comparison}.
This is expected, as the same CR injection power that had previously expanded throughout the sphere is now injected
in the denser region of the
flow (the torus region). Thus, the overall rate of p-p interactions is slightly increased and so the overall gamma-ray emission.
As a consequence, there is no relevant difference between the spherical and the torus-limited injections in the resulting SED profiles.

\section{Conclusions}
\label{sec:conclusions}
Point-like as well as diffuse VHE emission measured at the GC suggest the existence of a PeVatron in
the direction of SgrA*.
The site (or sites) of CR acceleration has not yet been determined by observations with the current IACTs. Future observations with the sensitivity of the forthcoming CTA may provide more localised VHE
fluxes able to trace the real sites of 
CR acceleration.

In this work, we investigate the RIAF of SgrA* as a potential PeVatron, assuming that CRs are accelerated by turbulent magnetic reconnection and we then derive the implied VHE emission fluxes produced exclusively
within the accretion flow plasma. 
Our approach is based on:
\begin{itemize}  
\setlength\itemsep{0em}
\item
numerical GRMHD plus GR radiative transfer, to model the RIAF environment of SgrA*, 
\item
CRs injection constrained by the magnetic reconnection power of the system which we calculate following
the model of \citep{2015ApJ...799L..20S}, and
\item
MC simulations of CRs propagation plus electromagnetic cascading, to calculate the VHE $\gamma$-ray
emission produced within the accretion flow environment.
\end{itemize}
We calculate the emission of CRs interacting with accretion flows considering three different mass accretion rates. 
Constraining the injection of CRs to the magnetic reconnection power, we find that systems 
with mass accretion rates $\gtrsim 10^{-7}$ M$_\odot$ yr$^{-1}$ produce VHE fluxes
which are compatible with the highest energy upper limits by H.E.S.S. and that could be 
potentially observed by CTA. 

For CR injection with power-law indices $\kappa=1.3-1.8$ and exponential cut-off energy $\epsilon_{cut}=0.5$
PeV (which are parameters consistent with particle acceleration by reconnection), such emission profiles
peak at $\sim 10^{13.5}$ eV and have no substantial contribution to the current data points of the 
source J1745-290 (see the curves of the models m$_{12}$, m$_{13}$, m$_{22}$ and m$_{23}$
in Fig.~\ref{VHE}). 

The neutrino fluxes associated to these emission profiles are negligible compared with the
IceCube neutrino diffuse emission.

The mass accretion rate threshold for detectable VHE fluxes derived here coincides with the upper limit
for the mass accretion rate of SgrA* inferred from rotation measures \citep{2007ApJ...654L..57M}.
Thus, if fluxes similar to  the emission profiles of models m$_{22}$ and m$_{23}$ (see Fig.~\ref{VHE})  
are detected, they might be correlated with episodic enhancements in the accretion rate of SgrA*.

\acknowledgments

We acknowledge partial support from the Brazilian agencies FAPESP (JCRR's grant: 2017/12188-5, RAB's grant: 2017/12828-4,  and EMGDP's grant: 2013/10559-5) and CNPq (306598/2009-4 grant). 
The simulations presented in this paper have made use of the computing facilities of the GAPAE group (FAPESP grant: 2013/10559-5, IAG-USP) and the Laboratory of Astroinformatics IAG/USP, NAT/Unicsul (FAPESP grant 2009/54006-4). 
\vspace{5mm}

\software{
{\tt harm} (\citealt{2003ApJ...589..444G}; \citealt{2006ApJ...641..626N}), 
{\tt grmonty} \citep{2009ApJS..184..387D},
{\tt CRPropa 3} \citep{2016JCAP...05..038A},
{\tt Matplotlib} \citep{Hunter:2007},
{\tt SciPy} \citep{numpy}
}.

\appendix

\section{Calculation of the VHE SED}
We calculate the energy flux of $\gamma$-rays within the energy bin $\epsilon$ measured at the
Earth as
\begin{equation}
\nu F_\nu = (4\pi R_s^2)^{-1}\epsilon^2 \frac{\dot{N}_\epsilon}{\Delta\epsilon},
\end{equation}
where $R_s=7.9$ kpc is the distance from SgrA* to us, $\Delta \epsilon$ is the size of the energy
bin, and $\dot{N}_\epsilon$ is the number of photons per unit time with energies between 
$\epsilon$ and $\epsilon+\Delta\epsilon$ that reach the spherical detection boundary
at $40 R_g$ (as a result of the hadronic interactions plus $\gamma$-$\gamma$/IC cascading simulated with the {\tt CRPropa 3} code, see Section 3).

As described in Section 3
(see also \citealt{2019arXiv190305249R}),
we emulate a continuous release of CRs by the RIAF of SgrA* 
during a time interval $\Delta t$
by injecting CRs at equally separated times $t_i$ (see Section 3) within the interval $\Delta t$ that contains four nearly steady state snapshots of the background accretion flow.
Thus, the rate of $\gamma$-rays leaving the 
detection boundary is calculated as 
\begin{equation}
\dot{N}_\epsilon =\frac{1}{\Delta t}
\sum\limits_{i=1}^{4} N_{\epsilon,i},
\end{equation} 
where $\Delta t$ is the time interval for photon detection (the same interval for CR injection),
and $N_{\epsilon,i}$
are the number of photons within the energy bin $\epsilon$, produced by CRs injected
at the time $t_i$. 

To save computational resources, we  inject $10^6$ CRs at each injection time $t_i$. This number of simulated CRs is
several orders of magnitude smaller than the number of
physical CRs expected in the physical system. Thus, we calculate the physical number of detected $\gamma$-rays,
$N_{\epsilon,i}$, assuming statistical convergence with the condition:
\begin{equation}
\label{ratio}
N_{\epsilon,i}=
\sum_{\epsilon_0}
\left(\frac{N_{CR,\epsilon_0,i}}{N_{CR,\epsilon_0,i}^{sim}} \right)
N^{sim}_{\epsilon(\epsilon_0),i},
\end{equation}
where the factor in the parenthesis is the ratio of physical to simulated CRs within the energy bin $\epsilon_0$, and 
$N_{\epsilon(\epsilon_0)}^{sim}$ is the number of photons in the output of the simulation within the energy bin $\epsilon$ that was ultimately
produced by CRs injected with initial
energy within the energy bin $\epsilon_0$.

The CR simulations are performed with power-law injection distribution of index 1. Thus to calculate
the flux of physical gamma rays corresponding to the injection of physical CRs with arbitrary power-law 
distribution $\kappa$, we normalise the distribution of the injected physical CRs
($N_{CR,\epsilon_0,i}\propto\epsilon_0^{-\kappa}\exp\left\{-\epsilon_0/\epsilon_{cut}\right\}$), with the
physical energy of CR injection $E_i$, and the distribution of simulated CRs ($N_{CR,\epsilon_0,i}^{sim}\propto \epsilon_0^{-1}$),
with the total number of simulated CRs $N_{CR}$. Then, the ratio in parenthesis in Eq.~\ref{ratio} is expressed in terms of the 
physical energy of CRs injection $E_i$, and the number of simulated CRs $N_{CR}$ as
\begin{equation}
\label{ratioEiNCR}
\frac{N_{CR,\epsilon_0,i}}{N^{sim}_{CR,\epsilon_0,i}} = 
\frac{E_i}{N_{CR,i}}
\frac{\ln (\epsilon_{max}/\epsilon_{min})}{\int_{\epsilon_{min}}^\infty
\epsilon^{1-\kappa}\exp\{-\epsilon/\epsilon_{cut}\} d\epsilon}
\epsilon_0^{1-\kappa}\exp\left\{
-\epsilon_0/\epsilon_{cut}
\right\}.
\end{equation}
For simplicity we consider the same amount of energy of the physical CRs in each injection time 
$E_i=W_{CR}\Delta t/4$, being $W_{CR}$ the power of physical CRs, assumed to be constant during the 
time interval $\Delta t$. 
Finally, combining Eqs.~\ref{ratio} and \ref{ratioEiNCR}, the rate of physical photons $\dot{N}_\epsilon$ 
can be expressed in terms of the power of CR injection, $W_{CR}$, as
\begin{equation}
\dot{N}_\epsilon= W_{CR} 
\frac{\ln \left(\epsilon_{max}/\epsilon_{min}\right) }{\int_{\epsilon_{min}}^\infty
\epsilon_0^{1-\kappa}\exp\{-\epsilon_0/\epsilon_{cut}\} d\epsilon_0}
\sum\limits_{i=1}^{4}\sum\limits_{\epsilon_0}
\frac{1}{N_{CR,i}}
\epsilon_0^{1-\kappa}
\exp\{-\epsilon_0/\epsilon_{cut}\}
N_{\epsilon(\epsilon_0),i}^{sim}\;.
\end{equation}

\section{Radial location of the production of VHE photons}

As described in Section 3, we use the gas density, magnetic and photon fields obtained 
with the GRMHD and GR leptonic radiative transfer simulations, to set the environment
in a flat spacetime where we perform CR and VHE photon propagation.
The error introduced by this
in the calculation of the overall VHE flux depends on the location where the photons are 
sourced.

\begin{figure}
   \centering
   \includegraphics[width=\columnwidth]{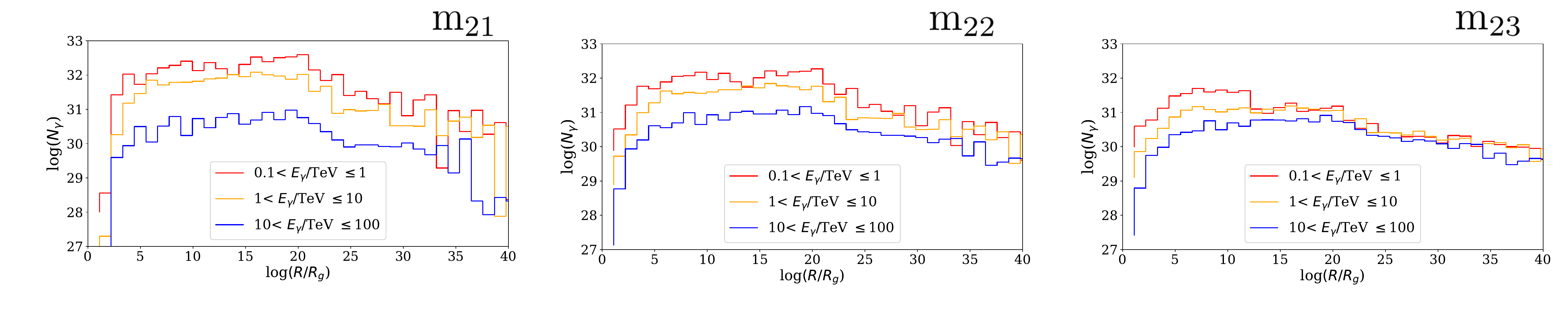}
      \caption{
      Distribution of photons as a function of their origin distance to the central BH, for different energy ranges. Each panel corresponds to the labelled model emission (see text).
}
         \label{R_origin}
\end{figure}

We show in Fig.~\ref{R_origin}, the  distribution of photons for different energy ranges,
as a function of the distance from the central BH where they were produced.
The distributions in these plots correspond 
to the emission models m$_{21}$, m$_{22}$, m$_{23}$ (the model parameters are in Table 1
of Section 4).
We can see that the origin of photons
is more uniformly distributed along the 40 $R_g$ volume for models m$_{22}$ and m$_{23}$ and for
photons with energy 10$<E/TeV<$100. The origin of photons with energies $<$1 TeV
 seems to be mostly sourced at distances $\lesssim 20R_g$ from the BH. 
 We then conclude that emission profiles peaking at energies $\gtrsim 10$ TeV like the emission
 profiles m$_{13}$ and m$_{23}$ (see Fig.~\ref{VHE}) appear to have the lowest error due to the neglection of
 the GR curvature effects
 in the calculation of the VHE SED.
 
\bibliographystyle{aasjournal}
\bibliography{refs} 

\end{document}